\documentclass[twocolumn,english,american]{revtex4-2}
\usepackage[T1]{fontenc}
\usepackage[utf8]{luainputenc}
\usepackage{geometry}
\usepackage{amsmath}
\usepackage{graphicx}

\makeatletter

\providecommand{\tabularnewline}{\\}

\makeatother

\usepackage{babel}
\begin{document}
\title{A Systematic Convergent Sequence of Approximations (of Integral Equation
Form) to the Solutions of the Hedin Equations }
\author{Garry Goldstein}
\address{garrygoldsteinwinnipeg@gmail.com}
\begin{abstract}
In many ways the solution to the Hedin equations represents an exact
solution to the many body problem. However, for most systems of practical
interest, the solution to the Hedin equations is rendered nearly numerically
intractable because the Hedin equations are of functional derivative
form. Integral equations, on the other hand, are much more numerically
tractable, then functional derivative equations, as they can often
be solved iteratively. In this work we present a systematic set of
integral equations (with no functional derivatives) - Hedin approximations
I, II, III, IV etc. - whose solutions converge to the solutions of
the exact Hedin equations. The Hedin approximations are well suited
to iterative numerical solutions (which we also describe). Furthermore
Hedin approximation I is just the GW approximation (as such this work
may be viewed as a systematic improvement of the GW approximation).
We present a systematic study of the Hedin equations for zero dimensional
field theory (which, in particular, is a method to enumerate Feynman
diagrams for field theories in arbitrary dimensions) and show better
and better convergence to the solutions of the Hedin equations for
higher and higher Hedin approximations, with Hedin approximations
I, II and III being explicitly studied. We, in particular, show that
the higher Hedin approximations capture more and and more Feynman
diagrams for the self energy. We also show that already Hedin approximation
II captures more diagrams then the state of the art diagrammatic vertex
corrections approach. Furthermore Hedin approximation III is a near
perfect match to the exact solutions of the Hedin equations, at least
in the zero dimensional case, and enumerates a large number of Feynman
diagrams.
\end{abstract}
\maketitle

\section{Introduction}\label{sec:Introduction}

It is known that Many Body Perturbation Theory (MBPT) \citep{Shavitt_2009,Stefanucci_2025,Aryasetiawan_2025}
is well suited for the study of complex quantum systems. MBPT, along
with its competitors - Quantum Monte Carlo (QMC) \citep{Blinder_2019,Becca_2017}
and Coupled Cluster (C.C.) calculations \citep{Shavitt_2009} - to
name a few, represents some of our best understanding of correlated
electron systems. One of the main advantages of MBPT is the direct
relation of the computed quantities, in MBPT, to physical observables.
In particular, single particle Green's functions are directly accessible
through MBPT - these can be used to compute density of states, as
well as the excitation spectra and, with the use of the Galitskii-Migdal
formula, the ground state energy of an interacting many-body system
\citep{Galitskii_1958,Aryasetiawan_2025,Aryasetiawan_2022}. Furthermore
two particle Green's functions, also accessible through MBPT, can
be used to compute linear response functions \citep{Aryasetiawan_2025,Kubo_1957}. 

One of the key methods within MBPT are the Hedin equations \citep{Hedin_1965,Lundqvist_1967,Lundqvist_1967(2),Aryasetiawan_2022,Aryasetiawan_2025}.
The Hedin equations are a relationship between the single particle
propagator $G$, the proper self energy $\Sigma$ (which does not
contain Hartree insertions), the effective potential $W$, the proper
polarization $P$, and the dressed vertex $\Gamma$. The Hedin equations
are capable of solving most many body problems in that they compute
the single particle Green's functions and, through the associated
Bethe-Salpeter equation, the two particle Green's functions \citep{Stefanucci_2025}.
Furthermore the Hedin equations are widely applicable, in particular
they apply when the many body Hamiltonian is of the form:
\begin{equation}
H=\sum_{i}h\left(\mathbf{x}_{i},\mathbf{p}_{i}\right)+\sum_{i<j}v\left(\mathbf{x}_{i},\mathbf{x}_{j}\right),\label{eq:Hamiltonian}
\end{equation}
spin orbit coupling and spin-spin interactions is also possible \citep{Aryasetiawan_2008,Lane_2025,Aryasetiawan_2009}
as well as phonons \citep{Stefanucci_2025} and transverse photons
\citep{Trevisanutto_2015}. Here $\mathbf{x}_{i}$ and $\mathbf{p}_{i}$
are the position and momentum. Here $h$ is the single particle part
while $v$ is the two particle part of the Hamiltonian, see however
also Refs. \citet{Yukalov_2016,Yukalov_2025}. The Hedin equations
for the Hamiltonian in Eq. (\ref{eq:Hamiltonian}) are given by \citep{Aryasetiawan_2022,Aryasetiawan_2025,Hedin_1965}: 
\begin{widetext}
\begin{align}
\Gamma\left(\mathbf{1},\mathbf{2},\mathbf{3}\right) & =\delta\left(\mathbf{1}-\mathbf{2}\right)\delta\left(\mathbf{2}-\mathbf{3}\right)+\int d\left(\mathbf{4},\mathbf{5},\mathbf{6},\mathbf{7}\right)\frac{\delta\Sigma\left(\mathbf{1},\mathbf{2}\right)}{\delta G\left(\mathbf{4},\mathbf{5}\right)}G\left(\mathbf{4},\mathbf{6}\right)\Gamma\left(\mathbf{6},\mathbf{7},\mathbf{3}\right)G\left(\mathbf{5},\mathbf{7}\right)\nonumber \\
P\left(\mathbf{1},\mathbf{2}\right) & =-i\int d\left(\mathbf{3},\mathbf{4}\right)G\left(\mathbf{1},\mathbf{3}\right)\Gamma\left(\mathbf{3},\mathbf{4},\mathbf{2}\right)G\left(\mathbf{4},\mathbf{1}\right)\nonumber \\
W\left(\mathbf{1},\mathbf{2}\right) & =v\left(\mathbf{1},\mathbf{2}\right)+\int d\left(\mathbf{3},\mathbf{4}\right)v\left(\mathbf{1},\mathbf{3}\right)P\left(\mathbf{3},\mathbf{4}\right)W\left(\mathbf{4},\mathbf{2}\right)\nonumber \\
\Sigma\left(\mathbf{1},\mathbf{2}\right) & =i\int d\left(\mathbf{3},\mathbf{4}\right)G\left(\mathbf{1},\mathbf{3}\right)W\left(\mathbf{4},1\right)\Gamma\left(\mathbf{3},\mathbf{2},\mathbf{4}\right)\nonumber \\
G\left(\mathbf{1},\mathbf{2}\right) & =g\left(\mathbf{1},\mathbf{2}\right)+\int d\left(\mathbf{3},\mathbf{4}\right)g\left(\mathbf{1},\mathbf{3}\right)\Sigma\left(\mathbf{3},\mathbf{4}\right)G\left(\mathbf{4},\mathbf{2}\right)\label{eq:Hedin}
\end{align}
\end{widetext}

Here $\mathbf{i}=\left(\mathbf{x}_{i},\sigma_{i},t_{i}\right)$ represents
the position, spin and time (both real and imaginary time is possible).
Keldysh indices may also be added \citep{Stefanucci_2025}. Furthermore,
$g\left(\mathbf{i},\mathbf{j}\right)$ is the bare single particle
Green's function with the inclusion of Hartree insertions. 

We explicitly see the functional derivative term $\frac{\delta\Sigma\left(\mathbf{1},\mathbf{2}\right)}{\delta G\left(\mathbf{4},\mathbf{5}\right)}$
in Eq. (\ref{eq:Hedin}). This term greatly increases the complexity
of the numerical solution to the Hedin equations \citep{Aryasetiawan_2022,Aryasetiawan_2025}.
In order to obtain equations without functional derivatives and reduce
the Hedin equations to integral equations (which are much easier to
solve then functional derivative equations) we propose to introduce
new independent variables, $\Gamma$, $P$, $W$, $\Sigma$, $G$,
$\frac{\delta}{\delta G}\Gamma$, $\frac{\delta}{\delta G}P$, $\frac{\delta}{\delta G}W$,
$\frac{\delta}{\delta G}\Sigma$, $\frac{\delta^{2}}{\delta G^{2}}\Gamma$,
$\frac{\delta^{2}}{\delta G^{2}}P$, $\frac{\delta^{2}}{\delta G^{2}}W$,
$\frac{\delta^{2}}{\delta G^{2}}\Sigma$, etc. (where, from now on,
we have dropped the labels $\mathbf{1}$, $\mathbf{2}$, $\mathbf{3}$
etc.) That is, we wish to have the functional derivatives become independent
variables. We then write the exact equations in terms of these variables
(that is use product rule for derivatives and differentiate under
the integral sign, repeatedly, on the r.h.s of Eq. (\ref{eq:Hedin}))
and expand the total number of equations to include those for the
functional derivatives - now promoted to independent variables - not
functional derivative relations. To make the calculations tractable
we truncate the equations at some order of derivative - that is drop
all terms of a higher derivative. Hedin approximation I (the GW approximation
\citep{Golze_2019,Marie_2023,Aryasetiawan_2025,Aryasetiawan_2022,Hedin_1965,Lundqvist_1967,Lundqvist_1967(2),Aryasetiawan_1997})
is when we truncated at zeroth order derivatives kept, Hedin approximation
II is when we truncate at first order derivatives kept, Hedin approximation
III is when we truncate at second order derivatives kept etc. Solutions
of these equations produces a sequence of better and better approximations
to the solutions of the Hedin equations with solutions to Hedin approximation
$n\rightarrow\infty$ converging to the exact results. Since Hedin
approximation I is just the GW approximation this work may be viewed
as a systematic improvement of the GW approximation \citep{Kutepov_2020,Aryasetiawan_2022,Aryasetiawan_2025,Hedin_1965,Lundqvist_1967,Lundqvist_1967(2),Aryasetiawan_1997}.
We further note that we do not explicitly compute any Feynman diagrams
in our approach (which is another method to improve GW approximation
\citep{Kutepov_2017,Kutepov_2017(2),Kutepov_2018,Schindlmayer_1997}),
this greatly simplifies the calculations.

We test this approach using zero dimensional field theory \citep{Molinari_2004,Pavlyukh_2007,Molinari_2006}.
It is known that the Hedin Equations in zero dimensions enumerate
Feynman diagrams for field theories in arbitrary dimensions \citep{Molinari_2004,Pavlyukh_2007,Molinari_2006}.
We show that Hedin approximations I, II, and III correspond to progressively
better and better enumerations with more and more diagrams kept. In
particular, already Hedin approximation II enumerates more Feynman
diagrams then the leading state of the art diagrammatic vertex corrections
approach currently used, as far as the author is aware \citep{Kutepov_2017,Kutepov_2017(2),Kutepov_2018,Schindlmayer_1997}.
Furthermore already Hedin approximation III enumerates most low order
Feynman diagrams and nearly matches the exact solutions of the Hedin
equations.

\section{Main Idea and Examples}\label{sec:Main-Idea}

The key idea, presented in this work, behind improving the GW approximation
to more accurately solve the Hedin equations, without the use of functional
derivatives or Feynman diagrams, is to introduce more independent
variables, that is promote $\frac{\delta}{\delta G}\Gamma$, $\frac{\delta}{\delta G}P$,
$\frac{\delta}{\delta G}W$, $\frac{\delta}{\delta G}\Sigma$, $\frac{\delta^{2}}{\delta G^{2}}\Gamma$,
$\frac{\delta^{2}}{\delta G^{2}}P$, $\frac{\delta^{2}}{\delta G^{2}}W$,
$\frac{\delta^{2}}{\delta G^{2}}\Sigma$, etc. to independent variables
(not functional derivative relations) and then write the exact equations
in terms of these variables (that is increase the number of integral
equations). This is done by repeated application of the product rule
for derivatives and differentiating under the integral sign of the
r.h.s. of Eq. (\ref{eq:Hedin}). One then truncates the equations
at some order of derivative with respect to $G$. If one truncates
at zero order derivatives one obtains Hedin approximation I (GW approximation)
if one truncates at first order derivatives one obtains Hedin approximation
II, if one truncates at second order derivatives one obtains Hedin
approximation III etc. We now present formulas for Hedin approximations
I, II, III explicitly and then we present the general formula for
Hedin approximation n. 

\subsection{ Hedin approximation I (GW Approximation)}\label{subsec:GW-Equations}

The Hedin equations for Hedin approximation I (GW approximation) reduce
to: 
\begin{align}
\Gamma & =1\nonumber \\
P & =GG\Gamma\nonumber \\
W & =v+vPW\nonumber \\
\Sigma & =GW\Gamma\nonumber \\
G & =g+g\Sigma G\label{eq:GWA}
\end{align}
Where, on top of dropping the labels $\mathbf{i},\mathbf{j}$....,
we drop, from now on, the integral signs and replace $\delta\left(\mathbf{1}-\mathbf{2}\right)\delta\left(\mathbf{2}-\mathbf{3}\right)\rightarrow1$.
These equations are well suited for iterative solutions as they from
a closed set of equations. Indeed one chooses some initial conditions
$\Gamma_{0},P_{0},W_{0},\Sigma_{0},G_{0}$ and writes the loop:
\begin{equation}
....\rightarrow\Gamma_{n}\rightarrow P_{n}\rightarrow W_{n}\rightarrow\Sigma_{n}\rightarrow G_{n}\rightarrow\Gamma_{n+1}\rightarrow...\label{eq:Cycle}
\end{equation}
Where one uses left to right assignments at each step e.g.
\begin{align}
\Gamma_{n} & =1\nonumber \\
P_{n} & =G_{n-1}G_{n-1}\Gamma_{n}\nonumber \\
W_{n} & =v+vP_{n}W_{n-1}\nonumber \\
\Sigma_{n} & =G_{n-1}W_{n}\Gamma_{n}\nonumber \\
G_{n} & =g+g\Sigma_{n}G_{n-1}\label{eq:GWA-1}
\end{align}
 We note that we have dropped the $\frac{\delta\Sigma}{\delta G}GG\Gamma$
\begin{equation}
\Gamma=1+\left(\frac{\delta\Sigma}{\delta G}GG\Gamma\right)\label{eq:Actual}
\end{equation}
That is truncated the equations at zero functional derivatives and
therefore had to add no new additional equations.

\subsection{Hedin approximation II}\label{subsec:Hedin-II}

We now write the equations for Hedin approximation II: 
\begin{align}
\Gamma & =1+\frac{\delta\Sigma}{\delta G}GG\Gamma\nonumber \\
P & =GG\Gamma\nonumber \\
W & =v+vPW\nonumber \\
\Sigma & =GW\Gamma\nonumber \\
G & =g+g\Sigma G\nonumber \\
\frac{\delta\Gamma}{\delta G} & =\frac{\delta}{\delta G}\left(1+\frac{\delta\Sigma}{\delta G}GG\Gamma\right)-\frac{\delta^{2}\Sigma}{\delta G^{2}}GG\Gamma\nonumber \\
 & =\frac{\delta\Sigma}{\delta G}GG\frac{\delta\Gamma}{\delta G}+\frac{\delta\Sigma}{\delta G}2G\Gamma\nonumber \\
\frac{\delta P}{\delta G} & =\frac{\delta}{\delta G}\left(GG\Gamma\right)=2G\Gamma+GG\frac{\delta\Gamma}{\delta G}\nonumber \\
\frac{\delta W}{\delta G} & =W\frac{\delta P}{\delta G}W\nonumber \\
\frac{\delta\Sigma}{\delta G} & =\frac{\delta}{\delta G}\left(GW\Gamma\right)=W\Gamma+G\frac{\delta W}{\delta G}\Gamma+GW\frac{\delta\Gamma}{\delta G}\label{eq:Hedin_II}
\end{align}
We note that we used the exact result that $\frac{\delta W}{\delta G}=W\frac{\delta P}{\delta G}W$
\citep{Stefanucci_2025} to simplify one of the equations. In Equation
(\ref{eq:Hedin_II}) we have dropped the term $\frac{\delta^{2}\Sigma}{\delta G^{2}}GG\Gamma$
from: 
\begin{align}
\frac{\delta\Gamma}{\delta G} & =\frac{\delta}{\delta G}\left(1+\frac{\delta\Sigma}{\delta G}GG\Gamma\right)\nonumber \\
 & =\frac{\delta\Sigma}{\delta G}GG\frac{\delta\Gamma}{\delta G}+\frac{\delta\Sigma}{\delta G}2G\Gamma+\left(\frac{\delta^{2}\Sigma}{\delta G^{2}}GG\Gamma\right)\label{eq:Dropped}
\end{align}
(which is obtained by applying the product rule on the r.h.s. of Eq.
(\ref{eq:Hedin}) and differentiating under the integral sign). That
is we truncated at one functional derivative kept. We note that all
the functional derivatives are to be viewed as independent variables
in Eq. (\ref{eq:Hedin_II}) and not as functional derivative relations.
As such we see there are now many new integral equations added to
the original Hedin equations - which are obtained by differentiating
the original Hedin equations and truncating at first derivative order.
With this, we now have the same number of equations as unknowns and
these equations are well suited for iterative solutions. Indeed one
writes: 
\begin{align}
.... & \rightarrow\Gamma_{n}\rightarrow P_{n}\rightarrow W_{n}\rightarrow\Sigma_{n}\rightarrow G_{n}\rightarrow\left(\frac{\delta\Gamma}{\delta G}\right)_{n}\rightarrow\nonumber \\
 & \rightarrow\left(\frac{\delta P}{\delta G}\right)_{n}\rightarrow\left(\frac{\delta W}{\delta G}\right)_{n}\rightarrow\left(\frac{\delta\Sigma}{\delta G}\right)_{n}\rightarrow\Gamma_{n+1}\rightarrow...\label{eq:Hedin_II_Cycle}
\end{align}
Where one uses right to left assignments at each step e.g. $\left(\frac{\delta\Gamma}{\delta G}\right)_{n}=\left(\frac{\delta\Sigma}{\delta G}\right)_{n-1}G_{n}G_{n}\left(\frac{\delta\Gamma}{\delta G}\right)_{n-1}+\left(\frac{\delta\Sigma}{\delta G}\right)_{n-1}2G_{n}\Gamma_{n}$. 

\subsection{Hedin approximation III}\label{subsec:Hedin-III}

For Hedin approximation III we write the equations:
\begin{align}
\Gamma & =1+\frac{\delta\Sigma}{\delta G}GG\Gamma\nonumber \\
P & =GG\Gamma\nonumber \\
W & =v+vPW\nonumber \\
\Sigma & =GW\Gamma\nonumber \\
G & =g+g\Sigma G\nonumber \\
\frac{\delta\Gamma}{\delta G} & =\frac{\delta\Sigma}{\delta G}GG\frac{\delta\Gamma}{\delta G}+\frac{\delta\Sigma}{\delta G}2G\Gamma+\frac{\delta^{2}\Sigma}{\delta G^{2}}GG\Gamma\nonumber \\
\frac{\delta P}{\delta G} & =2G\Gamma+GG\frac{\delta\Gamma}{\delta G}\nonumber \\
\frac{\delta W}{\delta G} & =W\frac{\delta P}{\delta G}W\nonumber \\
\frac{\delta\Sigma}{\delta G} & =W\Gamma+G\frac{\delta W}{\delta G}\Gamma+GW\frac{\delta\Gamma}{\delta G}\nonumber \\
\frac{\delta^{2}\Gamma}{\delta G^{2}} & =\frac{\delta^{2}\Sigma}{\delta G^{2}}GG\frac{\delta\Gamma}{\delta G}+\frac{\delta\Sigma}{\delta G}2G\frac{\delta\Gamma}{\delta G}+\frac{\delta\Sigma}{\delta G}GG\frac{\delta^{2}\Gamma}{\delta G^{2}}+\nonumber \\
 & +\frac{\delta^{2}\Sigma}{\delta G^{2}}2G\Gamma+\frac{\delta\Sigma}{\delta G}2\Gamma+\frac{\delta\Sigma}{\delta G}2G\frac{\delta\Gamma}{\delta G}+\nonumber \\
 & +\frac{\delta^{2}\Sigma}{\delta G^{2}}2G\Gamma+\frac{\delta^{2}\Sigma}{\delta G^{2}}GG\frac{\delta\Gamma}{\delta G}\nonumber \\
\frac{\delta^{2}P}{\delta G^{2}} & =2\Gamma+2G\frac{\delta\Gamma}{\delta G}+2G\frac{\delta\Gamma}{\delta G}+G^{2}\frac{\delta^{2}\Gamma}{\delta G^{2}}\nonumber \\
\frac{\delta^{2}W}{\delta G^{2}} & =W\frac{\delta P}{\delta G}W\frac{\delta P}{\delta G}W+W\frac{\delta^{2}P}{\delta G^{2}}W+W\frac{\delta P}{\delta G}W\frac{\delta P}{\delta G}W\nonumber \\
\frac{\delta^{2}\Sigma}{\delta G^{2}} & =\frac{\delta W}{\delta G}\Gamma+W\frac{\delta\Gamma}{\delta G}+\frac{\delta W}{\delta G}\Gamma+G\frac{\delta^{2}W}{\delta G^{2}}\Gamma+\nonumber \\
 & +G\frac{\delta W}{\delta G}\frac{\delta\Gamma}{\delta G}+W\frac{\delta\Gamma}{\delta G}+G\frac{\delta W}{\delta G}\frac{\delta\Gamma}{\delta G}+GW\frac{\delta^{2}\Gamma}{\delta G^{2}}\label{eq:Hedin_III}
\end{align}
Where we have dropped $\frac{\delta^{3}\Sigma}{\delta G^{3}}GG\Gamma$
from: 
\begin{align}
\frac{\delta^{2}\Gamma}{\delta G^{2}} & =\frac{\delta^{2}}{\delta G^{2}}\left(1+\frac{\delta\Sigma}{\delta G}GG\Gamma\right)\nonumber \\
 & =\frac{\delta^{2}\Sigma}{\delta G^{2}}GG\frac{\delta\Gamma}{\delta G}+\frac{\delta\Sigma}{\delta G}2G\frac{\delta\Gamma}{\delta G}+\frac{\delta\Sigma}{\delta G}GG\frac{\delta^{2}\Gamma}{\delta G^{2}}+\nonumber \\
 & +\frac{\delta^{2}\Sigma}{\delta G^{2}}2G\Gamma+\frac{\delta\Sigma}{\delta G}2\Gamma+\frac{\delta\Sigma}{\delta G}2G\frac{\delta\Gamma}{\delta G}+\frac{\delta^{2}\Sigma}{\delta G^{2}}2G\Gamma+\nonumber \\
 & +\frac{\delta^{2}\Sigma}{\delta G^{2}}GG\frac{\delta\Gamma}{\delta G}+\left(\frac{\delta^{3}\Sigma}{\delta G^{3}}GG\Gamma\right)\label{eq:Dropped_Hedin_III}
\end{align}
which would be obtained by using product rule on the r.h.s. of Eq.
(\ref{eq:Hedin}) and differentiating under the integral sign (that
is truncated to two derivatives kept). These equations are well suited
for iterative solutions as they from a closed set of equations as
we have added many additional equations beyond the Hedin ones - by
differentiating the Hedin equations and truncating at second derivative
order. To numerically solve these equations one writes: 
\begin{align}
.... & \rightarrow\Gamma_{n}\rightarrow P_{n}\rightarrow W_{n}\rightarrow\Sigma_{n}\rightarrow G_{n}\rightarrow\left(\frac{\delta\Gamma}{\delta G}\right)_{n}\rightarrow\nonumber \\
 & \rightarrow\left(\frac{\delta P}{\delta G}\right)_{n}\rightarrow\left(\frac{\delta W}{\delta G}\right)_{n}\rightarrow\left(\frac{\delta\Sigma}{\delta G}\right)_{n}\rightarrow\left(\frac{\delta^{2}\Gamma}{\delta G^{2}}\right)_{n}\rightarrow\nonumber \\
 & \rightarrow\left(\frac{\delta^{2}P}{\delta G^{2}}\right)_{n}\rightarrow\left(\frac{\delta^{2}W}{\delta G^{2}}\right)_{n}\rightarrow\left(\frac{\delta^{2}\Sigma}{\delta G^{2}}\right)_{n}\rightarrow\Gamma_{n+1}\rightarrow....\label{eq:Hedin_III_Cycle}
\end{align}
Where one uses right to left assignments at each step. Hedin IV and
higher are similar to Hedin I, II, III but more tedious. We will not
need their explicit forms below, but we given the general form of
Hedin n.

\subsection{Hedin approximation n}\label{subsec:Hedin-n}

For Hedin approximation n we write the equations: 
\begin{align}
\Gamma & =1+\frac{\delta\Sigma}{\delta G}GG\Gamma\nonumber \\
P & =GG\Gamma\nonumber \\
W & =v+vPW\nonumber \\
\Sigma & =GW\Gamma\nonumber \\
G & =g+g\Sigma G\nonumber \\
\cdots & =\cdots\nonumber \\
\cdots & =\cdots\nonumber \\
\cdots & =\cdots\nonumber \\
\cdots & =\cdots\nonumber \\
\frac{\delta^{n-1}\Gamma}{\delta G^{n-1}} & =\frac{\delta^{n-1}}{\delta G^{n-1}}\left(\frac{\delta\Sigma}{\delta G}GG\Gamma\right)-\frac{\delta^{n}\Sigma}{\delta G^{n}}GG\Gamma\nonumber \\
\frac{\delta^{n-1}P}{\delta G^{n-1}} & =\frac{\delta^{n-1}}{\delta G^{n-1}}\left(GG\Gamma\right)\nonumber \\
\frac{\delta^{n-1}W}{\delta G^{n-1}} & =\frac{\delta^{n-2}}{\delta G^{n-2}}\left(W\frac{\delta P}{\delta G}W\right)\nonumber \\
\frac{\delta^{n-1}\Sigma}{\delta G^{n-1}} & =\frac{\delta^{n-1}}{\delta G^{n-1}}\left(GW\Gamma\right)\label{eq:Hedin_n}
\end{align}
We notice that we have truncated the equations at $n-1$ functional
derivatives by dropping the term $\frac{\delta^{n}\Sigma}{\delta G^{n}}GG\Gamma$.
All the variables in Eq. (\ref{eq:Hedin_n}) are viewed as independent
variables and not as functional derivative relations. This makes these
equations well suited for iterative solutions as they are of integral
equation form. We note that for $n=I$ we explicitly have the GW equations
while for $n\rightarrow\infty$ the dropped term $\frac{\delta^{n}\Sigma}{\delta G^{n}}GG\Gamma$
has negligible effect on $\Gamma$, $P$, $W$, $\Sigma$ and $G$
so that the solutions of Hedin approximation $n\rightarrow\infty$
reduce to the solutions of the Hedin equations. We note that for finite
$n$, when we drop the term $\frac{\delta^{n}\Sigma}{\delta G^{n}}GG\Gamma$
- we drop a certain class of Feynman diagrams. As such, we do not
enumerate all Feynman diagrams that enter the MBPT Feynman diagram
series for any finite $n$ - which we emphasize is fully captured
by the exact Hedin Equations.

\section{Zero Dimensional Field Theory}\label{sec:Zero-Dimensional-Field}

Here we do some numerical simulations that confirm the efficiency
of our approach. We focus on zero dimensional field theory \citep{Molinari_2004,Pavlyukh_2007,Molinari_2006}
- which is an efficient way to enumerate Feynman diagrams for the
Schr$\ddot{\mathrm{o}}$dinger field theory in arbitrary dimensions
\citep{Molinari_2004,Fradkin_2021,Molinari_2006,Pavlyukh_2007}. We
explicitly calculate the number of Feynman diagrams that enter the
self energy at every order of perturbation theory upto fifth order
beyond Hartree-Fock (we focus on the self energy for simplicity and
concreteness; because of its physical significance \citep{Aryasetiawan_2025,Stefanucci_2025}
and to follow Ref. \citep{Molinari_2004}). We find that the exact
field theory captures more diagrams then Hedin approximation III,
which captures more diagrams the Hedin Approximation II which captures
more diagrams then the state of the art diagrammatic vertex corrections
approach which captures more diagrams then the GW approximation (Hedin
approximation I). We note that Hedin approximation III is already
a nearly perfect match to the exact solutions of the Hedin equations.
We note that the Hedin equations in zero dimensions are given by \citep{Molinari_2004}:
\begin{align}
\Gamma & =1+G^{2}\frac{\partial\Sigma}{\partial G}\Gamma\nonumber \\
P & =lG^{2}\Gamma\nonumber \\
W & =v+PW\nonumber \\
\Sigma & =GW\Gamma\nonumber \\
G & =g+g\Sigma G\label{eq:Heddin_zero}
\end{align}
Where we have introduced the fermion loop counting parameter $l$
following Ref. \citep{Molinari_2004}. 

\subsection{Diagrammatic vertex corrections to GW}\label{subsec:Iterating-diagrammatic-vertex}

We introduce the dimensionless interaction strength $x=g^{2}v$ following
Ref. \citep{Molinari_2004,Molinari_2006}. The Hedin equations for
the state of the art diagrammatic vertex correction approach \citep{Kutepov_2017,Kutepov_2017(2),Kutepov_2018}
can be written as:
\begin{align}
\Gamma_{D} & =1+\left(x\mathcal{G}_{D}^{2}\mathcal{W}_{D}+2lx^{2}\mathcal{G}_{D}^{4}\mathcal{W}_{D}^{2}\right)\Gamma_{D}\nonumber \\
\mathcal{P}_{D} & =\mathcal{G}_{D}^{2}\Gamma_{D}\nonumber \\
\mathcal{W}_{D} & =1+lx\mathcal{P}_{D}\mathcal{W}_{D}\nonumber \\
s_{D} & =\mathcal{G}_{D}\mathcal{W}_{D}\Gamma_{D}\nonumber \\
\mathcal{G}_{D} & =1+xs_{D}\mathcal{G}_{D}\label{eq:Diagrammatic_vertex}
\end{align}
Where we have introduced dimensionless parameters: $\Gamma_{D}=\Gamma,\mathcal{P}_{D}=\frac{P}{lg^{2}},\mathcal{W}_{D}=\frac{W}{v},s_{D}=\frac{\Sigma}{gv},\mathcal{G}_{D}=\frac{G}{g}$.
Iterating Eq. (\ref{eq:Diagrammatic_vertex}) starting at $\Gamma_{D}=1,\mathcal{P}_{D}=0,\mathcal{W}_{D}=1,s_{D}=0,\mathcal{G}_{D}=1$
with $l=1$ (the fermion loop counting parameter is set to one - so
we just count the total number of diagrams), we obtain that: 
\begin{equation}
s_{D}\left(x\right)=1+3x+16x^{2}+103x^{3}+733x^{4}+5556x^{5}+....\label{eq:s_D}
\end{equation}
We note that each coefficient in Eq. (\ref{eq:s_D}) counts the number
of diagrams of that order in the diagrammatic expansion of the self
energy captured by the diagrammatic vertex correction approach. That
is, the diagrammatic expansion in Eq. (\ref{eq:Diagrammatic_vertex})
captures one diagram at first oder of perturbation theory (note that
Hartree insertions are included in $g$ - so there is only one diagram
at first order in perturbation theory), three diagram at second order
perturbation theory, sixteen diagrams at third order perturbation
theory, 103 diagrams at fourth order perturbation theory, 733 diagrams
at fifth order perturbation theory and 5556 diagrams at sixth order
in perturbation theory in the bare couplings for the self energy.
This is true for arbitrary Hamiltonians of the form in Eq. (\ref{eq:Hamiltonian})
in arbitrary dimensions.

\subsection{Self energy for Hedin approximation II}\label{subsec:Iterating_Hedin_II}

Introducing the dimensionless variables: 
\begin{align}
 & \Gamma_{II}=\Gamma,\mathcal{P}_{II}=\frac{P}{lg^{2}},\mathcal{W}_{II}=\frac{W}{v},s_{II}=\frac{\Sigma}{gv},\mathcal{G}_{II}=\frac{G}{g},\nonumber \\
 & \delta\Gamma_{II}=g\frac{\partial\Gamma}{\partial G},\delta\mathcal{P}_{II}=\frac{\frac{\partial P}{\partial G}}{lg},\delta\mathcal{W}_{II}=\frac{g}{v}\frac{\partial W}{\partial G},\delta s_{II}=\frac{\frac{\partial\Sigma}{\partial G}}{v}\label{eq:Hedin_II_variables}
\end{align}
We get that the Hedin approximation II equations reduce to: 
\begin{align}
\Gamma_{II} & =1+x\delta s_{II}\mathcal{G}_{II}^{2}\Gamma_{II}\nonumber \\
\mathcal{P}_{II} & =\mathcal{G}_{II}^{2}\Gamma_{II}\nonumber \\
\mathcal{W}_{II} & =1+lx\mathcal{P}_{II}\mathcal{W}_{II}\nonumber \\
s_{II} & =\mathcal{G}_{II}\mathcal{W}_{II}\Gamma_{II}\nonumber \\
\mathcal{G}_{II} & =1+xs_{II}\mathcal{G}_{II}\nonumber \\
\delta\Gamma_{II} & =x\mathcal{G}_{II}^{2}\delta s_{II}\delta\Gamma_{II}+2x\delta s_{II}\mathcal{G}_{II}\Gamma_{II}\nonumber \\
\delta\mathcal{P}_{II} & =2\mathcal{G}_{II}\Gamma_{II}+l\mathcal{G}_{II}^{2}\delta\Gamma_{II}\nonumber \\
\delta\mathcal{W}_{II} & =lx\mathcal{W}_{II}^{2}\delta\mathcal{P}_{II}\nonumber \\
\delta s_{II} & =\mathcal{W}_{II}\Gamma_{II}+\mathcal{G}_{II}\delta\mathcal{W}_{II}\Gamma_{II}+\mathcal{G}_{II}\mathcal{W}_{II}\delta\Gamma_{II}\label{eq:Hedin_II_zero_dim}
\end{align}
We iterate Eq. (\ref{eq:Hedin_II_zero_dim}) starting with:
\begin{align}
 & \Gamma_{II}=1,\mathcal{P}_{II}=0,\mathcal{W}_{II}=1,s_{II}=0,\mathcal{G}_{II}=1,\nonumber \\
 & \delta\Gamma_{II}=0,\delta\mathcal{P}_{II}=0,\delta\mathcal{W}_{II}=0,\delta s_{II}=0\label{eq:Initial_Hedin_II}
\end{align}
We get that for $l=1$ we have that:
\begin{equation}
s_{II}\left(x\right)=1+3x+18x^{2}+146x^{3}+1385x^{4}+14344x^{5}+...\label{eq:Self_energy}
\end{equation}
We note that already Hedin II captures many more Feynman diagrams
then the leading order vertex correction scheme at every order in
perturbation theory (see also table \ref{tab:series}).

\subsection{Self energy within Hedin approximation III}\label{subsec:Iterating-Hedin-approximation}

Introducing the dimensionless variables: 
\begin{align}
 & \Gamma_{III}=\Gamma,\mathcal{P}_{III}=\frac{P}{lg^{2}},\mathcal{W}_{III}=\frac{W}{v},s_{III}=\frac{\Sigma}{gv},\mathcal{G}_{III}=\frac{G}{g},\nonumber \\
 & \delta\Gamma_{III}=g\frac{\partial\Gamma}{\partial G},\delta\mathcal{P}_{III}=\frac{\frac{\partial P}{\partial G}}{lg},\delta\mathcal{W}_{III}=\frac{g}{v}\frac{\partial W}{\partial G},\delta s_{III}=\frac{\frac{\partial\Sigma}{\partial G}}{v},\nonumber \\
 & \delta^{2}\Gamma_{III}=g^{2}\frac{\partial^{2}\Gamma}{\partial G^{2}},\delta\mathcal{P}_{III}=\frac{\frac{\partial^{2}P}{\partial G^{2}}}{l},\delta^{2}\mathcal{W}_{III}=\frac{g^{2}}{v}\frac{\partial^{2}W}{\partial G^{2}},\delta^{2}s_{III}=g\frac{\frac{\partial^{2}\Sigma}{\partial G^{2}}}{v}\label{eq:Hedin_III_variables}
\end{align}
then we have that the Hedin approximation III can be written as: 
\begin{widetext}
\begin{align}
\Gamma_{III} & =1+x\delta s_{III}\mathcal{G}_{III}^{2}\Gamma_{III}\nonumber \\
\mathcal{P}_{III} & =\mathcal{G}_{III}^{2}\Gamma_{III}\nonumber \\
\mathcal{W}_{III} & =1+lx\mathcal{P}_{III}\mathcal{W}_{III}\nonumber \\
s_{III} & =\mathcal{G}_{III}\mathcal{W}_{III}\Gamma_{III}\nonumber \\
\mathcal{G}_{III} & =1+xs_{III}\mathcal{G}_{III}\nonumber \\
\delta\Gamma_{III} & =x\mathcal{G}_{III}^{2}\delta s_{III}\delta\Gamma_{III}+2x\delta s_{III}\mathcal{G}_{III}\Gamma_{III}+x\delta^{2}s_{III}\mathcal{G}_{III}^{2}\Gamma_{III}\nonumber \\
\delta\mathcal{P}_{III} & =2\mathcal{G}_{III}\Gamma_{III}+l\mathcal{G}_{III}^{2}\delta\Gamma_{III}\nonumber \\
\delta\mathcal{W}_{III} & =lx\mathcal{W}_{III}^{2}\delta\mathcal{P}_{III}\nonumber \\
\delta s_{III} & =\mathcal{W}_{III}\Gamma_{III}+\mathcal{G}_{III}\delta\mathcal{W}_{III}\Gamma_{III}+\mathcal{G}_{III}\mathcal{W}_{III}\delta\Gamma_{III}\nonumber \\
\delta^{2}\Gamma_{III} & =2x\delta^{2}s_{III}\mathcal{G}_{III}^{2}\delta\Gamma_{III}+4x\delta s_{III}\mathcal{G}_{III}\delta\Gamma_{III}+x\delta s_{III}\mathcal{G}_{III}^{2}\delta^{2}\Gamma_{III}+4x\delta^{2}s_{III}\mathcal{G}_{III}\Gamma_{III}+2x\delta s_{III}\Gamma_{III}\nonumber \\
\delta^{2}\mathcal{P}_{III} & =2\Gamma_{III}+4\mathcal{G}_{III}\delta\Gamma_{III}+\mathcal{G}_{III}^{2}\delta^{2}\Gamma_{III}\nonumber \\
\delta^{2}\mathcal{W}_{III} & =2l^{2}x\mathcal{W}_{III}^{3}\delta\mathcal{P}_{III}^{2}+lx\mathcal{W}_{III}^{2}\delta^{2}\mathcal{P}_{III}\nonumber \\
\delta^{2}s_{III} & =2\delta\mathcal{W}_{III}\Gamma_{III}+2\mathcal{W}_{III}\delta\Gamma_{III}+2\mathcal{G}_{III}\delta\mathcal{W}_{III}\delta\Gamma_{III}+\mathcal{G}_{III}\delta^{2}\mathcal{W}_{III}\Gamma_{III}+\mathcal{G}_{III}\mathcal{W}_{III}\delta^{2}\Gamma_{III}\label{eq:Hedin_III_zero_dim}
\end{align}
\end{widetext}

Whereby we obtain for $l=1$: 
\begin{equation}
s_{III}\left(x\right)=1+3x+20x^{2}+186x^{3}+2153x^{4}+29024x^{5}+...\label{eq:Self_energy-3}
\end{equation}
We shall see below that this is a nearly perfect match to the exact
series with nearly all Feynman diagrams captured (see also table \ref{tab:series}).

\subsection{Final results}\label{subsec:Final-results}

\selectlanguage{english}%
\begin{table}
\caption{A table of the power series expansions (enumeration of the number
of Feynman diagrams) of the dimensionless self energy for various
methods: GW, diagrammatic vertex corrections, Hedin approximation
II, Hedin approximation III and the exact solutions.}\label{tab:series}

\begin{tabular}{|c|c|}
\hline 
Method & \foreignlanguage{american}{Self Energy Series}\tabularnewline
\hline 
\hline 
GW & \foreignlanguage{american}{$1+2x+7x^{2}+30x^{3}+143x^{4}+728x^{5}+....$}\tabularnewline
\hline 
\foreignlanguage{american}{Vertex} & \foreignlanguage{american}{$1+3x+16x^{2}+103x^{3}+733x^{4}+5556x^{5}+....$}\tabularnewline
\hline 
\foreignlanguage{american}{Hedin II} & \foreignlanguage{american}{$1+3x+18x^{2}+146x^{3}+1385x^{4}+14344x^{5}+...$}\tabularnewline
\hline 
\foreignlanguage{american}{Hedin III} & \foreignlanguage{american}{$1+3x+20x^{2}+186x^{3}+2153x^{4}+29024x^{5}+...$}\tabularnewline
\hline 
\foreignlanguage{american}{Exact} & \foreignlanguage{american}{$1+3x+20x^{2}+189x^{3}+2232x^{4}+31130x^{5}+...$}\tabularnewline
\hline 
\end{tabular}
\end{table}

\selectlanguage{american}%
We note that the exact series in $x$ for the Hedin equations for
the self energy is given by \citep{Molinari_2004}: 
\begin{equation}
s_{E}\left(x\right)=1+3x+20x^{2}+189x^{3}+2232x^{4}+31130x^{5}+...\label{eq:Self_energy-1}
\end{equation}
We now emphasize that Hedin approximation III captures one out of
one leading order diagrams, three out of three second order diagrams,
twenty out of twenty third order diagrams, 186 out of 189 fourth order
diagrams, 2153 out of 2232 fifth order diagrams, 29024 out of 31130
sixth order diagrams for the self energy and as such is a near perfect
match to the exact results. 

\begin{figure}
\begin{centering}
\includegraphics[width=1\columnwidth]{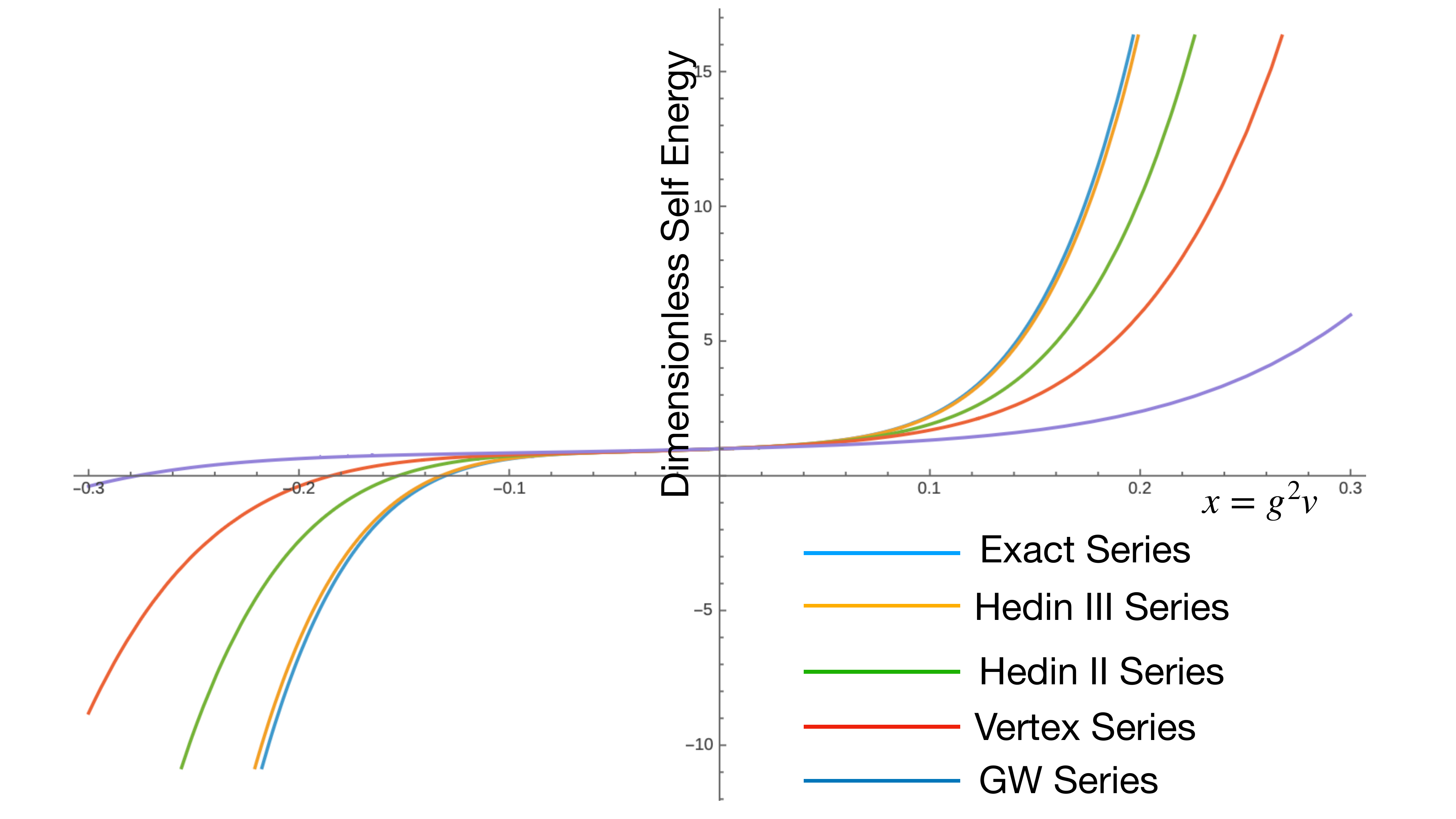}
\par\end{centering}
\caption{Self energy graphs for the GW series, diagrammatic vertex correction,
Hedin approximation II, Hedin approximation III and the exact series.
We notice that Hedin approximation II is already better (captures
the self energy more closely) then the state of the art diagrammatic
expansion. Furthermore Hedin III is nearly on top of the exact self
energy in zero dimensions and captures nearly all Feynman diagrams.}\label{fig:Hedin_graph}
\end{figure}

We note that the GW series (Hedin approximation I) is given by \citep{Molinari_2004}:
\begin{equation}
s_{GW}\left(x\right)=1+2x+7x^{2}+30x^{3}+143x^{4}+728x^{5}+...\label{eq:Self_energy-2}
\end{equation}
We see that Hedin II captures many more diagrams then the GW approximation
this is further supported by the curves Fig. (\ref{fig:Hedin_graph})
and the series in table \ref{tab:series}). We note again that Hedin
II is already sufficient to beat the state of the art vertex corrections
as it captures many more Feynman diagrams, see Fig (\ref{fig:Hedin_graph}).
Furthermore as we see in Fig. (\ref{fig:Hedin_graph}) the Hedin III
approximation is a near perfect match to the exact self energy curves
(see Table \ref{tab:series} and Fig. (\ref{fig:Hedin_graph})).

\selectlanguage{english}%

\section{Conclusions and Discussions}\label{sec:Conclusions}

In this work we have studied a sequence of integral equation approximations
to the solutions of the Hedin equations: Hedin approximation I, II,
III, IV etc. We have shown that Hedin approximation I is equivalent
to the GW approximation while the solutions to Hedin approximations
$n\rightarrow\infty$ are equivalent to the exact solutions to the
Hedin equations (as such we have introduced a systematic method to
improve the GW approximation without resorting to functional derivatives
or Feynman diagrams). The main idea for this sequence of approximations
was to promote functional derivatives (with respect to the propagator
$G$) to independent variables, study the exact set of equations thereby
obtained (that is differentiate the original Hedin equations, in Eq.
(\ref{eq:Hedin}), using product rule inside the integral sign and
generate many new equations) and truncate at some order of derivative
kept (that is drop all terms with higher derivatives in $G$). This
leads to purely integral equations with no functional derivatives
- which are known to be significantly easier to solve. We presented
the example of zero dimensional field theory and showed explicitly
for Hedin approximations I, II and III that higher Hedin approximations
are closer to the exact solutions of the Hedin equations and retain
more terms in a power series expansion in the dimensionless interaction
strength. We also showed that Hedin approximation II already captures
more Feynman diagrams then the state of the art diagrammatic vertex
correction calculation \citep{Kutepov_2017,Kutepov_2017(2),Schindlmayer_1997},
while Hedin III is nearly a perfect match to the exact Hedin equations. 

In the future it would be of interest to extend these calculations
to the uniform electron gas, Hubbard model, simple compounds and elements
as well as small molecules. For small molecules the GW approximation
(Hedin approximation I) is known to work very well \citep{Bruneval_2021}
so it would be of great interest to see the improvement of the higher
Hedin approximations: II, III, IV etc. It would also be of interest
to combine Hedin approximations II, III, IV etc. with Dynamical Mean
Field Theory (DMFT) much like GW+eDMFT \citep{Kotliar_2006}. It would
also be of interest to extend these ideas to phonons as in the Hedin-Baym
equations \citep{Stefanucci_2025} and to spin-orbit coupling \foreignlanguage{american}{\citep{Aryasetiawan_2008,Lane_2025,Aryasetiawan_2009}.}

\selectlanguage{american}%
We would like to further note that in this work we are proposing a
new method to solve systems of differential equations. Indeed suppose
we have a differential equation of the form: 
\begin{equation}
\mathbf{0}=\mathbf{F}\left(\mathbf{x},\mathbf{X},\frac{\partial\mathbf{X}_{i}}{\partial\mathbf{x}_{j}}\right)\label{eq:Equation}
\end{equation}
That is $\mathbf{X}$ is a vector field over $\mathbf{x}$ that satisfies
the differential equation in Eq. (\ref{eq:Equation}) for some vector
field $\mathbf{F}$. Higher derivative versions are also possible.
We can simplify this equation by writing a series of approximations
of the form: 
\begin{equation}
\mathbf{0}=\mathbf{F}\left(\mathbf{x},\mathbf{X},\frac{\partial\mathbf{X}_{i}}{\partial\mathbf{x}_{j}}\rightarrow0\right)\label{eq:Equation_I}
\end{equation}
and call it approximation I, and of the form
\begin{align}
\mathbf{0} & =\mathbf{F}\left(\mathbf{x},\mathbf{X},\frac{\partial\mathbf{X}_{i}}{\partial\mathbf{x}_{j}}\right)\nonumber \\
\mathbf{0} & =\frac{\partial\mathbf{F}_{i}}{\partial\mathbf{x}_{j}}\left(\mathbf{x},\mathbf{X},\frac{\partial\mathbf{X}_{k}}{\partial\mathbf{x}_{l}},\frac{\partial^{2}\mathbf{X}_{m}}{\partial\mathbf{x}_{j}\partial\mathbf{x}_{l}}\rightarrow0\right)\label{eq:Equation_II}
\end{align}
and call it approximation II, and of the form: 
\begin{align}
\mathbf{0} & =\mathbf{F}\left(\mathbf{x},\mathbf{X},\frac{\partial\mathbf{X}_{i}}{\partial\mathbf{x}_{j}}\right)\nonumber \\
\mathbf{0} & =\frac{\partial\mathbf{F}_{i}}{\partial\mathbf{x}_{j}}\left(\mathbf{x},\mathbf{X},\frac{\partial\mathbf{X}_{k}}{\partial\mathbf{x}_{l}},\frac{\partial^{2}\mathbf{X}_{m}}{\partial\mathbf{x}_{j}\partial\mathbf{x}_{l}}\right)\nonumber \\
\mathbf{0} & =\frac{\partial^{2}\mathbf{F}_{i}}{\partial\mathbf{x}_{j}\partial\mathbf{x}_{k}}\left(\mathbf{x},\mathbf{X},\frac{\partial\mathbf{X}_{p}}{\partial\mathbf{x}_{l}},\frac{\partial^{2}\mathbf{X}_{m}}{\partial\mathbf{x}_{p}\partial\mathbf{x}_{l}},\frac{\partial^{3}\mathbf{X}_{m}}{\partial\mathbf{x}_{j}\partial\mathbf{x}_{k}\partial\mathbf{x}_{l}}\rightarrow0\right)\label{eq:Equation_III}
\end{align}
which is approximation III and so forth. In the future it would be
of interest to see applications of this technique to the solutions
of differential or functional derivative equations, in particular
to see which equations besides the Hedin ones are particularly amenable
to this technique. Based on our experience with Hedin equations these
should be singular differential equations - where the highest derivative
terms are multiplied by functions that vanish at specific points and
therefore contribute little to the equations (may be dropped to leading
order near those points).

\selectlanguage{english}%
\textbf{Acknowledgements: }The author would like to acknowledge Gabriel
Kotliar for many useful discussions.

\appendix

\section{Some variations}\label{sec:Some-variations}

\subsection{First variation}\label{subsec:First-variation}
\selectlanguage{american}%

\subsubsection{Hedin Equations}\label{subsec:Hedin-Equations}

\selectlanguage{english}%
The Hedin Equations may also be written as \citep{Molinari_2004}:
\selectlanguage{american}%
\begin{widetext}
\begin{align}
\Gamma\left(\mathbf{1},\mathbf{2},\mathbf{3}\right) & =\delta\left(\mathbf{1}-\mathbf{2}\right)\delta\left(\mathbf{2}-\mathbf{3}\right)+\int d\left(\mathbf{4},\mathbf{5},\mathbf{6},\mathbf{7}\right)\frac{\delta\Sigma\left(\mathbf{1},\mathbf{2}\right)}{\delta g\left(\mathbf{4},\mathbf{5}\right)}g\left(\mathbf{4},\mathbf{3}\right)g\left(\mathbf{5},\mathbf{3}\right)\nonumber \\
P\left(\mathbf{1},\mathbf{2}\right) & =-i\int d\left(\mathbf{3},\mathbf{4}\right)G\left(\mathbf{1},\mathbf{3}\right)\Gamma\left(\mathbf{3},\mathbf{4},\mathbf{2}\right)G\left(\mathbf{4},\mathbf{1}\right)\nonumber \\
W\left(\mathbf{1},\mathbf{2}\right) & =v\left(\mathbf{1},\mathbf{2}\right)+\int d\left(\mathbf{3},\mathbf{4}\right)v\left(\mathbf{1},\mathbf{3}\right)P\left(\mathbf{3},\mathbf{4}\right)W\left(\mathbf{4},\mathbf{2}\right)\nonumber \\
\Sigma\left(\mathbf{1},\mathbf{2}\right) & =i\int d\left(\mathbf{3},\mathbf{4}\right)G\left(\mathbf{1},\mathbf{3}\right)W\left(\mathbf{4},1\right)\Gamma\left(\mathbf{3},\mathbf{2},\mathbf{4}\right)\nonumber \\
G\left(\mathbf{1},\mathbf{2}\right) & =g\left(\mathbf{1},\mathbf{2}\right)+\int d\left(\mathbf{3},\mathbf{4}\right)g\left(\mathbf{1},\mathbf{3}\right)\Sigma\left(\mathbf{3},\mathbf{4}\right)G\left(\mathbf{4},\mathbf{2}\right)\label{eq:Hedin_variant_I}
\end{align}
\end{widetext}

This version of the Hedin equations is also amenable to our approximations
method we call these the Hedin approximations I', II', III' etc.

\subsubsection{Hedin approximations n'}\label{subsec:Hedin-approximations-n'}

The Hedin equations for the Hedin approximation n' are given by:
\begin{align}
\Gamma & =1+\frac{\delta\Sigma}{\delta g}gg\nonumber \\
P & =GG\Gamma\nonumber \\
W & =v+vPW\nonumber \\
\Sigma & =GW\Gamma\nonumber \\
G & =g+g\Sigma G\nonumber \\
\cdots & =\cdots\nonumber \\
\cdots & =\cdots\nonumber \\
\cdots & =\cdots\nonumber \\
\cdots & =\cdots\nonumber \\
\frac{\delta^{n'-1}\Gamma}{\delta g^{n'-1}} & =\frac{\delta^{n'-1}}{\delta g^{n'-1}}\left(\frac{\delta\Sigma}{\delta g}gg\right)-\frac{\delta^{n'}\Sigma}{\delta g^{n'}}gg\nonumber \\
\frac{\delta^{n'-1}P}{\delta g^{n'-1}} & =\frac{\delta^{n'-1}}{\delta g^{n'-1}}\left(GG\Gamma\right)\nonumber \\
\frac{\delta^{n'-1}W}{\delta g^{n'-1}} & =\frac{\delta^{n'-2}}{\delta g^{n'-2}}\left(W\frac{\delta P}{\delta g}W\right)\nonumber \\
\frac{\delta^{n'-1}\Sigma}{\delta g^{n'-1}} & =\frac{\delta^{n'-1}}{\delta g^{n'-1}}\left(GW\Gamma\right)\nonumber \\
\frac{\delta^{n'-1}G}{\delta g^{n'-1}} & =\frac{\delta^{n'-1}}{\delta g^{n'-1}}\left(g+g\Sigma G\right)\label{eq:Hedin_n'}
\end{align}
Which is very similar to Hedin approximation n, in particular we have
that Hedin approximation $n'\rightarrow\infty$ converges to the exact
Hedin equations. These equations may be efficiently iterated in order
to obtain their solutions.

\subsection{Second variation}\label{subsec:Second-variation}

\subsubsection{Hedin Equations}\label{subsec:Hedin-Equations-1}

The Hedin equations may also be written as:
\begin{widetext}
\begin{align}
\Gamma\left(\mathbf{1},\mathbf{2}\mid\mathbf{3},\mathbf{4}\right) & =\delta\left(\mathbf{1}-\mathbf{3}\right)\delta\left(\mathbf{2}-\mathbf{4}\right)+\int d\left(\mathbf{5},\mathbf{6}\right)\frac{\delta\Sigma\left(\mathbf{1},\mathbf{2}\right)}{\delta G\left(\mathbf{5},\mathbf{6}\right)}K\left(\mathbf{5},\mathbf{6}\mid\mathbf{3},\mathbf{4}\right)\nonumber \\
K\left(\mathbf{1},\mathbf{2}\mid\mathbf{3},\mathbf{4}\right) & =i\int d\left(\mathbf{5},\mathbf{6}\right)\left[G\left(\mathbf{1},\mathbf{5}\right)G\left(\mathbf{6},\mathbf{2}\right)\right]\Gamma\left(\mathbf{5},\mathbf{6}\mid\mathbf{3},\mathbf{4}\right)\nonumber \\
W\left(\mathbf{1},\mathbf{2}\mid\mathbf{3},\mathbf{4}\right) & =v\left(\mathbf{1},\mathbf{2}\mid\mathbf{3},\mathbf{4}\right)+\int d\left(\mathbf{5},\mathbf{6},\mathbf{7},\mathbf{8}\right)v\left(\mathbf{1},\mathbf{2}\mid\mathbf{5},\mathbf{6}\right)K\left(\mathbf{5},\mathbf{6}\mid\mathbf{7},\mathbf{8}\right)W\left(\mathbf{7},\mathbf{8}\mid\mathbf{3},\mathbf{4}\right)\nonumber \\
\Sigma\left(\mathbf{1},\mathbf{2}\right) & =i\int d\left(\mathbf{3},\mathbf{4},\mathbf{5},\mathbf{6}\right)G\left(\mathbf{3},\mathbf{4}\right)\Gamma\left(\mathbf{4},\mathbf{2}\mid\mathbf{5},\mathbf{6}\right)W\left(\mathbf{5},\mathbf{6}\mid\mathbf{3},\mathbf{1}\right)\nonumber \\
G\left(\mathbf{1},\mathbf{2}\right) & =g\left(\mathbf{1},\mathbf{2}\right)+\int d\left(\mathbf{3},\mathbf{4}\right)g\left(\mathbf{1},\mathbf{3}\right)\Sigma\left(\mathbf{3},\mathbf{4}\right)G\left(\mathbf{4},\mathbf{2}\right)\label{eq:Hedin_variant_II}
\end{align}

This version of the Hedin equations is also amenable to our approximations
method we call these the Hedin approximations I'', II'', III''
etc. 
\end{widetext}

\subsubsection{Hedin approximations n''}\label{subsec:Hedin-approximations-n}

The Hedin equations for the Hedin approximation n'' are given by:

\begin{align}
\Gamma & =1+\frac{\delta\Sigma}{\delta G}K\nonumber \\
K & =GG\Gamma\nonumber \\
W & =v+vKW\nonumber \\
\Sigma & =G\Gamma W\nonumber \\
G & =g+g\Sigma G\nonumber \\
\cdots & =\cdots\nonumber \\
\cdots & =\cdots\nonumber \\
\cdots & =\cdots\nonumber \\
\cdots & =\cdots\nonumber \\
\frac{\delta^{n"-1}\Gamma}{\delta G^{n"-1}} & =\frac{\delta^{n"-1}}{\delta G^{n"-1}}\left(\frac{\delta\Sigma}{\delta G}K\right)-\frac{\delta^{n"}\Sigma}{\delta G^{n"}}K\nonumber \\
\frac{\delta^{n"-1}K}{\delta G^{n"-1}} & =\frac{\delta^{n"-1}}{\delta G^{n"-1}}\left(GG\Gamma\right)\nonumber \\
\frac{\delta^{n"-1}W}{\delta G^{n"-1}} & =\frac{\delta^{n"-1}}{\delta G^{n"-1}}\left(v+vKW\right)\nonumber \\
\frac{\delta^{n"-1}\Sigma}{\delta G^{n"-1}} & =\frac{\delta^{n"-1}}{\delta G^{n"-1}}\left(GW\Gamma\right)\label{eq:Hedin_n-1}
\end{align}
We note that $v$ and $G$ are independent constants therefore $\frac{\delta v}{\delta G}=0$.
This slight variation is similar to the equations in the main text,
in particular we have that Hedin approximation $n"\rightarrow\infty$
converges to the exact Hedin equations.

\end{document}